\documentstyle[11pt]{article}
\begin{document}
\renewcommand{ \baselinestretch} {2}

\centerline{\Large\bf A realization of Yangian and its
applications to the bi-spin system} \centerline{\Large\bf in an
external magnetic field} \vspace{1.0cm} \centerline{ Shuo
Jin$^{a,b,}$\footnote{ Corresponding author. {\sl Fax:
}+86-22-235-01532; {\sl E-mail address: }jinshuo@eyou.com}, Kang
Xue$^c$, Bing-Hao Xie$^{a,b}$} \vspace{0.4cm} \centerline{\it a
Theoretical Physics Division, Nankai Institute of Mathematics,}
\centerline{\it Nankai University, Tianjin 300071, P.R.china}
\centerline{\it b Liuhui Center for Applied Mathematics,}
\centerline{\it Nankai University and Tianjin University, Tianjin
300071, P.R.China} \centerline{\it c Physics Department, Northeast
Normal University,} \centerline{\it Changchun, Jilin 130024,
P.R.china} 

\vspace{2.0cm} \large\noindent{\bf Abstract}
\\

Yangian $Y(sl(2))$ is realized in the bi-spin system coupled with
a time-dependent external magnetic field. It is shown that
$Y(sl(2))$ generators can describe the transitions between the
``spin triplet'' and the ``spin singlet'' that evolve with time.
Furthermore, new transition operators between the states with
Berry phase factor and those between the states of Nuclear
Magnetic Resonance (NMR) are presented.

\vspace{3.5cm} \noindent {\sl PACS:
}02.20.-a, 03.65.-w


\vspace{3cm}

\pagebreak


\section*{1. Introduction }
\vspace{0.2cm}
 \ \ \ \
 Yangian Algebras were established by  Drinfeld \cite{1,2} based on the investigation of
 Yang-Baxter equation. In recent years, many
 works in studying Yangian and its applications have been made,
 including the Yangian symmetry in quantum integrable models (such as
 Haldane-Shastry model \cite{3},
 Calogero-Sutherland model \cite{4}, and Hubbard model \cite{5,6}) and the realization of Yangian in quantum mechanics \cite{7,8,9}.
 In quantum mechanics, the Yangian associated  with $sl(2)$ called $Y(sl(2))$
 has been realized in angular momentum quantum mechanics systems \cite{7}, hydrogen
 atom \cite{8,9} and other systems.The sense of Yangian generators' transition operator has also been known. This is natural
 because Yangian algebras belong to hopf algebras
 and regard Lie algebras as their subalgebras.

 It was pointed out in \cite{7} that, $Y(sl(2))$ can be constructed
 in a bi-spin system
 with spin $\hat S^1$ and $\hat S^2$ ($\hat S^k$ is the $k$th spin operator).
 It has been known that the Yangian generators
 $\{\hat I,\hat J\}$ can realize the transitions between the spin
 triplet and the spin singlet, which are the bases of quantum states of the bi-spin
 system (each $S^k$ has spin-1/2)
 that does not evolve with time.
 An interesting problem arises that if this
 system evolves with time, such as is coupled with a time-dependent external magnetic field,
 what will happen for the transition
 operators.
 A physical picture of the transition between the states is useful to help us
  understanding this time-dependent problem.
 In this paper, we will consider this issue. First, we will
 illustrate that $Y(sl(2))$ generators $\{\hat {\mathcal I}(t),\hat {\mathcal J}(t)\}$ still
 play the role of transition operators.
 Then, we discuss the transitions between the states with Berry phase
 factor of this system under the adiabatic condition. And finally, using the combination of the
 $Y(sl(2))$ generators $\{\hat {\mathcal I}(t),\hat {\mathcal J}(t)\}$, the transition
 operators between the Nuclear Magnetic Resonance (NMR) states are also obtained.

\vspace{0.2cm}
\section*{{\noindent \bf 2. Realization of $Y(sl(2))$ and transition operators for the bi-spin system in an external magnetic field
}}
\vspace{0.2cm}

 \ \ \ \ $Y(sl(2))$ is formed by a set
of operators \{$\hat{I},\hat{J}$\} obeying the commutation
relations \cite{1,7}:
\begin{eqnarray}
\label{definition} &&[\hat I_\alpha ,\hat I_\beta ] =i\epsilon
_{\alpha \beta\gamma }\hat I_{\gamma
},\nonumber \\
&&[\hat I_\alpha ,\hat J_\beta ]=i\epsilon _{\alpha \beta\gamma
}\hat J_\gamma \;\;\;\;\;\;
(\alpha,\beta,\gamma=1,2,3),\nonumber \\
&&[\hat J_{\pm },[\hat J_3,\hat J_{\pm }]] =\hat I_{\pm}(\hat
J_{\pm
}\hat I_3-\hat I_{\pm}\hat J_3), \nonumber\\
&&[\hat J_3,[\hat J_{+},\hat J_{-}]]=\hat I_3(\hat I_{+}\hat
J_{-}-\hat J_{+}\hat I_{-}).
\end{eqnarray}
$\hat{I}$ stands for the generators of $sl(2)$. Hereafter for any
operators $\hat A_{\pm}=\hat A_{1}{\pm}i\hat A_{2}$
($\hat{A}=\hat{I},\hat{J}$) are understood. In a bi-spin system
with spin $\hat S^1$ and $\hat S^2$, the $Y(sl(2))$ generators
take the form \cite{7}:
\begin{eqnarray}
\label{realization1} &&\hat{I}\equiv\hat{S}=\hat{S}^1+\hat{S}^2,
\nonumber
\\
&&\hat{J}={\mu}(\hat{S}^1-\hat{S}^2)+\frac{ih}{2}\hat{S}^1\times\hat{S}^2
\end{eqnarray}
where $\mu$ and $h$ are arbitrary parameters. In this paper, we
consider the case that the
 two spins are all equal to 1/2. Then direct calculation shows that the Yangian
generators $\hat J_{\alpha}\;(\alpha=\pm,3)$ describe the
transitions between the spin triplet ($S=1$)
\begin{equation}
\label{triplet1} \mid
X_{11}\rangle=\mid\uparrow\uparrow\rangle,\;\;\mid
X_{10}\rangle=\frac{1}{\sqrt2}(\mid\uparrow\downarrow\rangle+\mid
\downarrow\uparrow\rangle),\;\;\mid
X_{1-1}\rangle=\mid\downarrow\downarrow\rangle
\end{equation}
and the spin singlet ($S=0$)
\begin{equation}
\label{singlet1} \mid
X_{00}\rangle=\frac{1}{\sqrt2}(\mid\uparrow\downarrow\rangle-\mid\downarrow\uparrow\rangle).
\end{equation}
The transition relations are:
\begin{eqnarray}
\label{transition1}
&&\hat J_+\mid X_{1-1}\rangle=\sqrt2(\mu-\frac{h}{4})\mid X_{00}\rangle, \nonumber \\
&&\hat J_3\mid X_{10}\rangle=(\mu-\frac{h}{4})\mid X_{00}\rangle, \nonumber\\
&&\hat J_-\mid X_{11}\rangle=-\sqrt2(\mu-\frac{h}{4})\mid X_{00}\rangle, \nonumber \\
&&\hat J_+\mid X_{00}\rangle=-\sqrt2(\mu+\frac{h}{4})\mid X_{11}\rangle, \nonumber \\
&&\hat J_3\mid X_{00}\rangle=(\mu+\frac{h}{4})\mid X_{10}\rangle, \nonumber \\
&&\hat J_-\mid X_{00}\rangle=\sqrt2(\mu+\frac{h}{4})\mid
X_{1-1}\rangle.
\end{eqnarray}
Obviously the Lie algebra generators $\hat I_\pm$ can only
describe the transitions between the spin triplet, but $\hat
J_{\alpha}\ (\alpha=\pm,3)$ can realize the transitions between
the states with different Lie-algebra weights.

The Hamiltonian of the bi-spin system coupled with a
time-dependent external field $\bf B(t)$ reads
\begin{equation}
\label{h2} \hat {\mathcal
H}(t)=-\frac{1}{2}\hat{S}^1\cdot\hat{S}^2-\gamma{\bf{B}}(t)\cdot\hat{S}
\end{equation}
where $\gamma$ is gyromagnetic ratio. To study this system, we
firstly write a simple hamiltonian describing the bi-spin system
in a steady magnetic field as follows:
\begin{equation}
\label{h1} \hat H=-\frac{1}{2}\hat{S}^1\cdot\hat{S}^2-g\hat S_3
\end{equation}
where $g$ is a constant. It is easy to see that the eigenstates of
$\hat H$ are still the spin singlet and the spin triplet. So
Yangian generators $\{\hat I,\hat J\}$ can work well in this
system. An interesting picture will appear if the maguitude of the
magnetic field $\mid {\bf B}(t)\mid$ is chosen to be a
time-independent constant ${\bf B}_0$. Comparing Eq.(\ref{h1})
with Eq.(\ref{h2}), we find that through introducing the following
unitary transformation operator
\begin{eqnarray}
\label{u}
&&\hat U(t)=\prod^{2}_{i=1}\hat U^i(t),\nonumber\\
&&\hat U^i(t)=\left(\frac{2}{{\bf B}_0+B_3(t){\bf
B}_0}\right)^{1/2}{\bf{G}}(t)\cdot\hat{S}^i\;\;\;\;\;(i=1,2)
\end{eqnarray}
where
\begin{equation}
\label{g} {\bf{G}}(t)=\left(B_1(t),B_2(t),B_3(t)+{\bf B}_0\right),
\end{equation}
there exists the following transformation relation:
\begin{equation}
\label{hrelation} \hat {\mathcal H}(t)=\hat U(t)\hat H\hat
U^{-1}(t)
\end{equation}
when we choose $g=-\gamma{\bf B}_0$.

From Eq.(\ref{hrelation}), it is shown that the eigenvalues of
$\hat {\mathcal H}(t)$ are the same as those of $\hat H$, and the
eigenstate $\mid{\mathcal X}_{jm}(t)>$ of $\hat {\mathcal H}(t)$
can be got from the transformation of the eigenstate
$\mid{X}_{jm}>$ of $\hat H$, i.e.,
\begin{equation}
\label{xrelation} \mid{\mathcal X}_{jm}(t)>=\hat
U(t)\mid{X}_{jm}>\;\;\;\;\;(jm=11,10,1-1,00)
\end{equation}
where $\mid{X}_{jm}>$ is the spin triplet (Eqs.(\ref{triplet1}))
or the spin singlet (Eq.(\ref{singlet1})).

Through the unitary transformation $\hat U(t)$, the spin
realization of $Y(sl(2))$ has the time-dependent generators
\begin{eqnarray}
\label{realization2}
&&\hat{\mathcal I}(t)=\hat U(t)\hat{I}\hat U^{-1}(t),\nonumber\\
&&\hat{\mathcal J}(t)=\hat U(t)\hat{J}\hat U^{-1}(t),
\end{eqnarray}
which can be verified to still satisfy the definition
Eqs.(\ref{definition}) of $Y(sl(2))$. $\hat {\mathcal
J}_{\alpha}(t)\;(\alpha=\pm,3)$ are the transition operators
between the``spin triplet'' ($S=1$) $\{\mid{\mathcal
X}_{11}(t)>,\mid{\mathcal X}_{10}(t)>,\mid{\mathcal
X}_{1-1}(t)>\}$ and the ``spin singlet'' ($S=0$) $\mid{\mathcal
X}_{00}(t)>$ in a time-dependent magnetic field, and the
transition relations take the same form as
Eqs.(\ref{transition1}):
\begin{eqnarray}
\label{transition2}
&&\hat {\mathcal J}_+(t)\mid{\mathcal X}_{1-1}(t)>=\sqrt2(\mu-\frac{h}{4})\mid{\mathcal X}_{00}(t)>,\nonumber\\
&&\hat {\mathcal J}_3(t)\mid{\mathcal X}_{10}(t)>=(\mu-\frac{h}{4})\mid{\mathcal X}_{00}(t)>,\nonumber\\
&&\hat {\mathcal J}_-(t)\mid{\mathcal X}_{11}(t)>=-\sqrt2(\mu-\frac{h}{4})\mid{\mathcal X}_{00}(t)>,\nonumber\\
&&\hat {\mathcal J}_+(t)\mid{\mathcal X}_{00}(t)>=-\sqrt2(\mu+\frac{h}{4})\mid{\mathcal X}_{11}(t)>,\nonumber\\
&&\hat {\mathcal J}_3(t)\mid{\mathcal X}_{00}(t)>=(\mu+\frac{h}{4})\mid{\mathcal X}_{10}(t)>,\nonumber\\
&&\hat {\mathcal J}_-(t)\mid{{\mathcal
X}}_{00}(t)>=\sqrt2(\mu+\frac{h}{4})\mid{\mathcal X}_{1-1}(t)>.
\end{eqnarray}

In fact, the generators $\hat{{\mathcal I}}(t)$ and
$\hat{{\mathcal J}}(t)$ vary with $\hat U(t)$ because the
selection of $\hat U(t)$ is not exclusive. The choice of $\hat
U(t)$ does not affect the action of generators of Yangian
$Y(sl(2))$, so we can choose the unitary operator $\hat U(t)$ as
simple as possible. Taking advantages of the above results, we
will give two physical applications in the following.

\vspace{0.2cm}
\section*{{\noindent \bf 3. Transition operators between the states with Berry phase factor}}
\vspace{0.2cm}

 \ \ \ \ It has been shown that Berry
phase \cite{10} play a fundamental and an important role in
quantum mechanics in the past two decades. Berry phase can be
verified by experiments \cite{11}, and it can not be neglected in
many physics lessons. Recently, Berry phase of the bi-spin system
coupled with an external magnetic field has been studied
\cite{12,13}. Now we will give the transition operators between
the states with Berry phase factor.

Consider the bi-spin system in a rotating magnetic field described
by
\begin{equation}
\label{b}
B_1(t)=B_0\sin\theta\cos\omega_0t,\;\;\;\;B_2(t)=-B_0\sin\theta\sin\omega_0t,\;\;\;\;B_3(t)=B_0\cos\theta
\end{equation}
where constant $B_0$ is the strength (as referred to before) of
the magnetic field. Substituting Eqs.(\ref{b}) into Eqs.(\ref{u})
and Eq.(\ref{xrelation}), we can give the eigenstates of
Hamiltonian Eq.(\ref{h2}) that is the ``triplet'' ($S=1$)
\begin{eqnarray}
\label{triplet2}
&&\mid{\mathcal X}_{11}(t)>=\frac{1}{2}(1+\cos\theta)\mid{X}_{11}>+\frac{1}{\sqrt2}e^{-i\omega_0t}\sin\theta\mid{X}_{10}>+\frac{1}{2}(1-\cos\theta)e^{-2i\omega_0t}\mid{X}_{1-1}>,\nonumber\\
&&\mid{\mathcal X}_{10}(t)>=\frac{1}{\sqrt2}\sin\theta e^{i\omega_0t}\mid{X}_{11}>-\cos\theta\mid{X}_{10}>-\frac{1}{\sqrt2}\sin\theta e^{-i\omega_0t}\mid{X}_{1-1}>,\nonumber\\
&&\mid{\mathcal X}_{1-1}(t)>=\frac{1}{2}(1-\cos\theta)e^{2i\omega_0t}\mid{X}_{11}>-\frac{1}{\sqrt2}e^{i\omega_0t}\sin\theta\mid{X}_{10}>+\frac{1}{2}(1+\cos\theta)\mid{X}_{1-1}>,\nonumber\\
\end{eqnarray}
and the ``singlet'' ($S=0$)
\begin{equation}
\label{singlet2} \mid{\mathcal X}_{00}(t)>=-\mid{X}_{00}>.
\end{equation}
Under the adiabatic condition, the state with Berry phase factor
has the form:
\begin{equation}
\label{psi}
\mid\psi_{jm}(t)>=\exp{\{-\frac{i}{\hbar}E_{jm}t\}}\exp{\{i\gamma_{jm}(t)\}}\mid{\mathcal
X}_{jm}(t)>
\end{equation}
where $E_{jm}$ is the eigenvalue of Hamiltonian Eq.(\ref{h2}) and
is given by
\begin{equation}
\label{e1} E_{11}=-\frac{1}{8}-\gamma
B_0,\;\;\;E_{10}=-\frac{1}{8},\;\;\;E_{1-1}=-\frac{1}{8}+\gamma
B_0,\;\;\;E_{00}=-\frac{3}{8}.
\end{equation}
On the other hand, the Berry phase $\gamma_{jm}(t)$ reads
\begin{equation}
\label{p}
\gamma_{11}(t)=\omega_0(1-\cos\theta)t,\;\;\;\gamma_{1-1}(t)=-\omega_0(1-\cos\theta)t,\;\;\;\gamma_{10}(t)=\gamma_{00}(t)=0.
\end{equation}

By comparing Eqs.(\ref{transition2}) with Eq.(\ref{psi}), we
immediately find that
\begin{eqnarray}
\label{transition3}
&&\hat {\mathcal J}_+(t)\mid\psi_{1-1}(t)>=\sqrt2(\mu-\frac{h}{4})\exp{\{-\frac{i}{\hbar}(E_{1-1}-E_{00})t\}}\exp{\{i\gamma_{1-1}(t)\}}\mid\psi_{00}(t)>,\nonumber\\
&&\hat {\mathcal J}_3(t)\mid\psi_{10}(t)>=(\mu-\frac{h}{4})\exp{\{-\frac{i}{\hbar}(E_{10}-E_{00})t\}}\mid\psi_{00}(t)>,\nonumber\\
&&\hat {\mathcal J}_-(t)\mid\psi_{11}(t)>=-\sqrt2(\mu-\frac{h}{4})\exp{\{-\frac{i}{\hbar}(E_{11}-E_{00})t\}}\exp{\{i\gamma_{11}(t)\}}\mid\psi_{00}(t)>,\nonumber\\
&&\hat {\mathcal J}_+(t)\mid\psi_{00}(t)>=-\sqrt2(\mu-\frac{h}{4})\exp{\{-\frac{i}{\hbar}(E_{11}-E_{00})t\}}\exp{\{-i\gamma_{1-1}(t)\}}\mid\psi_{11}(t)>,\nonumber\\
&&\hat {\mathcal J}_3(t)\mid\psi_{00}(t)>=(\mu+\frac{h}{4})\exp{\{\frac{i}{\hbar}(E_{10}-E_{00})t\}}\mid\psi_{10}(t)>,\nonumber\\
&&\hat {\mathcal
J}_-(t)\mid\psi_{00}(t)>=\sqrt2(\mu+\frac{h}{4})\exp{\{\frac{i}{\hbar}(E_{1-1}-E_{00})t\}}\exp{\{-i\gamma_{1-1}(t)\}}\mid\psi_{1-1}(t)>.
\end{eqnarray}
We have solved the transition problems between the states with
Berry phase factor by applying $\hat {\mathcal
J}_\alpha(t)\;(\alpha=\pm,3)$.

\vspace{0.2cm}
\section*{{ \noindent \bf 4. Transition operators between the states of NMR}
}
\vspace{0.2cm}

 \ \ \ \ NMR is a very important
experiment technique based on  quantum mechanics \cite{14,15}. It
has been made rapid progress since 1945. Very recently, NMR is
used to realize the Geometric Quantum Computation \cite{16,17}.
The motivation for this section is to find the transition
operators between the NMR states. We  choose the same magnetic
field and the Hamiltonian as that in the former sections.

By solving the $Schr\ddot{o}dinger$ equation and utilizing the
magnetic resonance condition (MNC)
\begin{equation}
\label{mnc}
\omega_0=\gamma B_3
\end{equation}
we can get the states of NMR by choosing different initial states.
These states are the time-dependent combination of the eigenstates
of the Hamiltonian Eq.(\ref{h2}). The ``triplet'' ($S=1$) has the
forms:
\begin{eqnarray}
\label{triplet3} &&\mid\phi_{11}(t)>=a_1(t)\mid {\mathcal
X}_{11}(t)>+a_2(t)\mid {\mathcal X}_{10}(t)>+a_3(t)\mid
{\mathcal X}_{1-1}(t)>,\nonumber\\
&&\mid\phi_{10}(t)>=b_1(t)\mid {\mathcal X}_{11}(t)>+b_2(t)\mid
{\mathcal X}_{10}(t)>+b_3(t)\mid
{\mathcal X}_{1-1}(t)>,\nonumber\\
&&\mid\phi_{1-1}(t)>=c_1(t)\mid {\mathcal X}_{11}(t)>+c_2(t)\mid
{\mathcal X}_{10}(t)>+c_3(t)\mid {\mathcal X}_{1-1}(t)>.
\end{eqnarray}
In the process of calculating, the eigenvalues of deformed wave
functions under MNC Eq.(\ref{mnc}) have the exact values:
\begin{equation}
\label{e2} E_{11}'=-\frac{1}{8}-\gamma
B_1,\;\;\;E_{10}'=-\frac{1}{8},\;\;\;E_{1-1}'=-\frac{1}{8}+\gamma
B_1,\;\;\;E_{00}'=-\frac{3}{8}.
\end{equation}
 The time-dependent coefficients in Eqs.(\ref{triplet3}) are
\begin{eqnarray}
\label{abc}
a_1(t)&=&\frac{1}{2}\exp{\{-\frac{i}{\hbar}E_{10}'t\}}\exp{\{i\omega_0t\}}\nonumber\\
&&(\cos\theta\cos\omega_0t+\cos\omega_0t\cos\omega_1t+i\sin\omega_0t+i\cos\theta\sin\omega_0t\cos\omega_1t+i\sin\theta\sin\omega_1t),\nonumber\\
a_2(t)&=&\frac{1}{\sqrt2}\exp{\{-\frac{i}{\hbar}E_{10}'t\}}(\sin\theta\cos\omega_0t+i\sin\theta\sin\omega_0t\cos\omega_1t-i\cos\theta\sin\omega_1t),\nonumber\\
a_3(t)&=&-\frac{1}{2}\exp{\{-\frac{i}{\hbar}E_{10}'t\}}\exp{\{-i\omega_0t\}}\nonumber\\
&&(\cos\theta\cos\omega_0t-\cos\omega_0t\cos\omega_1t-i\sin\omega_0t+i\cos\theta\sin\omega_0t\cos\omega_1t+i\sin\theta\sin\omega_1t),\nonumber\\
b_1(t)&=&\frac{1}{\sqrt2}\exp{\{-\frac{i}{\hbar}E_{10}'t\}}\exp{\{i\omega_0t\}}(i\cos\omega_0t\sin\omega_1t-\cos\theta\sin\omega_0t\sin\omega_1t+\sin\theta\cos\omega_1t),\nonumber\\
b_2(t)&=&-\exp{\{-\frac{i}{\hbar}E_{10}'t\}}(\sin\theta\sin\omega_0t\sin\omega_1t+\cos\theta\cos\omega_1t),\nonumber\\
b_3(t)&=&\frac{1}{\sqrt2}\exp{\{-\frac{i}{\hbar}E_{10}'t\}}\exp{\{-i\omega_0t\}}(i\cos\omega_0t\sin\omega_1t+\cos\theta\sin\omega_0t\sin\omega_1t-\sin\theta\cos\omega_1t),\nonumber\\
c_1(t)&=&\frac{1}{2}\exp{\{-\frac{i}{\hbar}E_{10}'t\}}\exp{\{i\omega_0t\}}\nonumber\\
&&(-\cos\theta\cos\omega_0t+\cos\omega_0t\cos\omega_1t+-i\sin\omega_0t+i\cos\theta\sin\omega_0t\cos\omega_1t+i\sin\theta\sin\omega_1t),\nonumber\\
c_2(t)&=&\frac{1}{\sqrt2}\exp{\{-\frac{i}{\hbar}E_{10}'t\}}(\sin\theta\cos\omega_0t-i\sin\theta\sin\omega_0t\cos\omega_1t+i\cos\theta\sin\omega_1t),\nonumber\\
c_3(t)&=&-\frac{1}{2}\exp{\{-\frac{i}{\hbar}E_{10}'t\}}\exp{\{-i\omega_0t\}}\nonumber\\
&&(-\cos\theta\cos\omega_0t-\cos\omega_0t\cos\omega_1t+i\sin\omega_0t+i\cos\theta\sin\omega_0t\cos\omega_1t+i\sin\theta\sin\omega_1t)\nonumber\\
\end{eqnarray}
where $\omega_1=\gamma B_1$. The ``singlet'' ($S=0$) is
\begin{equation}
\label{singlet3} \mid\phi_{00}(t)>=-\exp{\{-iE_{00}'t\}}\mid
{\mathcal X}_{00}(t)>.
\end{equation}
From Eqs.(\ref{transition2}), Eqs.(\ref{triplet3}) and
Eqs.(\ref{singlet3}), after a rather long calculation we get the
following relations:
\begin{eqnarray}
\label{transition4}
&&\hat {\mathcal J}_-(t)\mid\phi_{11}(t)>=-\sqrt2a_1(t)(\mu-\frac{h}{4})exp{\{iE_{00}'t\}}\mid\phi_{00}(t)>,\nonumber\\
&&\{a_3(t)\hat {\mathcal J}_-(t)-a_1(t)\hat {\mathcal J}_+(t)+\sqrt2a_2(t)\hat {\mathcal J}_3(t)\}\mid\phi_{00}(t)>=\sqrt2(\mu+\frac{h}{4})\exp{\{-iE_{00}'t\}}\mid\phi_{11}(t)>,\nonumber\\
&&\hat {\mathcal J}_3(t)\mid\phi_{10}(t)>=b_2(t)(\mu-\frac{h}{4})exp{\{iE_{00}'t\}}\mid\phi_{00}(t)>,\nonumber\\
&&\{b_3(t)\hat {\mathcal J}_-(t)-b_1(t)\hat {\mathcal J}_+(t)+\sqrt2b_2(t)\hat {\mathcal J}_3(t)\}\mid\phi_{00}(t)>=\sqrt2(\mu+\frac{h}{4})\exp{\{-iE_{00}'t\}}\mid\phi_{10}(t)>,\nonumber\\
&&\hat {\mathcal J}_+(t)\mid\phi_{1-1}(t)>=\sqrt2c_3(t)(\mu-\frac{h}{4})exp{\{iE_{00}'t\}}\mid\phi_{00}(t)>,\nonumber\\
&&\{c_3(t)\hat {\mathcal J}_-(t)-c_1(t)\hat {\mathcal J}_+(t)+\sqrt2c_2(t)\hat {\mathcal J}_3(t)\}\mid\phi_{00}(t)>=\sqrt2(\mu+\frac{h}{4})\exp{\{-iE_{00}'t\}}\mid\phi_{1-1}(t)>.\nonumber\\
\end{eqnarray}

From Eqs.(\ref{transition4}), we can draw the conclusion that
$\hat {\mathcal J}_\alpha(t)\;(\alpha=\pm,3)$ and its combination
construct the transition operators between the NMR states (for
simplicity, among the above relations, we have chosen the
constants which can be modified to zero or 1 if it is possible).

\vspace{2.0cm} \section*{\noindent
\bf 5.Conclusions} \vspace{0.2cm}

\ \ \ \ In summary, we get a
time-dependent realization of $Y(sl(2))$ in the bi-spin system
coupled with a time-dependent external magnetic field. Although we
can verify that $Y(sl(2))$ does not describe the symmetry of the
system which we study because  $\hat {\mathcal J}(t)$ does not
commute with $\hat{\mathcal H}(t)$, we concentre on the transition
function of Yangian. We find that the generators $\{\hat {\mathcal
I}(t),\hat {\mathcal J}(t)\}$ of $Y(sl(2))$ can describe a new
picture of transition between two quantum states at any time. For
briefness, we have neglected the part of transitions that are
described by the Lie algebra operators. As far as we know, many
interesting investigations have relations with the bi-spin model
coupled with a time-dependent external magnetic field, such as
 Geometric Quantum Computation \cite{16,17}, entanglement \cite{18}. At last, we emphasis that
  Yangian algebras belong to hopf algebras and take Lie algebras
 as their subalgebras, so the Yangian operators  can connect
 the physical states with different Lie-algebra weights.
 It's reasonable to believe that the more interesting
 Yangian realization and
the more useful physical applications should be found.
\vspace{0.3cm}
\section*{\noindent \bf Acknowledegement} \vspace{0.2cm}
 \ \ \ \ We thank Dr. Hong-Biao Zhang
and Dr. Hui Jing for valuable discussions. This work is supported
by NSF of China.

\vspace{3mm}

\end{document}